\documentclass[
preprint,
aps,prd,
nofootinbib,
superscriptaddress,
showpacs,
tightenlines,
]{revtex4}
\usepackage{amsmath}
\usepackage{amssymb}
\usepackage{bm}
\usepackage[dvips]{color,graphicx}

\begin{document}

\title{Generalized Parton Distributions from Hadronic Observables: Non-Zero Skewness}

\author{Saeed Ahmad}
\email[E-mail: ]{sa8y@virginia.edu}
\affiliation{University of Virginia, 382 McCormick Road, Charlottesville, Virginia 22904, USA.}

\author{Heli Honkanen} 
\email[E-mail: ]{hh9e@virginia.edu}
\affiliation{University of Virginia, 382 McCormick Road, Charlottesville, Virginia 22904, USA.}

\author{Simonetta Liuti} 
\email[E-mail: ]{sl4y@virginia.edu}
\affiliation{University of Virginia, 382 McCormick Road, Charlottesville, Virginia 22904, USA.}

\author{Swadhin~K.~Taneja} 
\email[E-mail: ]{taneja@cpht.polytechnique.fr}
\affiliation{Ecole Polytechnique, CPHT, F91128 Palaiseau Cedex, France}


\begin{abstract}
We propose a physically motivated parametrization for the unpolarized 
generalized parton distributions, $H$ and $E$, valid at both zero and non-zero values
of the skewness variable, $\zeta$. 
Our approach follows a previous detailed study of the $\zeta=0$ case
where $H$ and $E$ were determined using constraints from simultaneous fits 
of the experimental data on both 
the nucleon elastic form factors and the deep inelastic structure functions
in the non singlet sector. 
Additional constraints at $\zeta \neq 0$ are provided by lattice 
calculations of the higher moments of generalized parton distributions.
We illustrate a method for extracting generalized parton distributions from 
lattice moments based on a reconstruction using 
sets of orthogonal polynomials.
The inclusion in our fit of data on Deeply Virtual Compton Scattering is also discussed.
Our method provides a step towards an extraction of generalized distributions 
based on a global fit of the available data within the given set of constraints. 
\end{abstract}

\maketitle

\baselineskip 3.5ex
\section{Introduction}
A completely new perspective was given to the study of hadronic structure, with
the observation that deep inelastic exclusive experiments, 
such as Deeply Virtual Compton Scattering (DVCS), and hard Exclusive Meson Production (EMP),
could in principle  allow one to access spatial configurations of partons
besides the relatively well known longitudinal momentum fraction ones \cite{Bur}. 
Because of the specific role played by the partons' spatial coordinates, 
one can also envisage progress towards the solution of the proton spin crisis through studies of
the orbital angular momentum contribution of different partonic components \cite{Ji1}.    
Information from exclusive deep inelastic scattering  
is coded in terms of Generalized Parton Distributions (GPDs) representing the soft
matrix elements in the process.
Since they were first introduced  \cite{DMul1,Ji1,Rad1}, much progress has been made 
in determining their general properties
in terms of the relevant kinematical variables, including the longitudinal momentum fraction 
taken by the struck quark,
$X$, the four-momentum transfer defining the scale of the deep inelastic process, 
$Q^2 \equiv -q_\mu^2$, 
the four-momentum transfer squared between the initial 
and final proton states, $t \equiv \Delta^2$, and the longitudinal momentum
transfer fraction of the initial proton momentum, 
the ``skewness'', $\zeta$, or $\xi$,
(for reviews see Refs.\cite{Die_rev,BelRad}). 

At present, a central issue is the definition of a quantitative, reliable 
approach beyond the construction of GPDs from specific models and/or particular 
limiting cases, that can incorporate incoming experimental data 
in a variety of ranges of $Q^2$ and $\Delta \equiv (\zeta,t)$, 
as they gradually become available. 
One hopes eventually to obtain a similar sophistication 
level as for the Parton Distribution
Functions (PDFs) extracted from inclusive Deep Inelastic Scattering (DIS).   
We address this issue in the present paper. 

The matching between measured quantities and perturbative QCD (PQCD) based predictions   
for DVCS/EMP proceeds, in principle, 
similarly to the inclusive case, with a few important caveats due
to the fact that 
GPDs describe {\em amplitudes} and are therefore much more elusive observables
from the practical point of view of experimental measurements, than the PDFs.
In DIS the cross sections at large enough
$Q^2$ and $s$ (the total invariant mass), where a partonic picture is expected to hold, measure 
directly the Bjorken $x$ dependent structure functions; $x$ is interpreted as the 
momentum fraction carried by the partons, modulo target mass corrections. A widely used approach
is ``global fitting'' whereby PDFs are obtained from experimental data by assuming a parametric 
functional form at a given scale, which is: {\it i)} evolved to the scale of the data; 
{\it ii)} convoluted with the appropriate perturbatively calculated coefficient functions; 
{\it iii)} compared to the data, thus determining the parameters. 

In DVCS/EMP similar factorization theorems as for the inclusive case exist. However, 
the phase space that has to be covered in order to extract GPDs has two extra dimensions, 
$\zeta$ and $t$, 
and the cross sections
are written in terms of convolutions of GPDs over the variable $X$. 
Experiments delivering sufficiently accurate data have just begun. 
The comparison with experiment and the formulation of parametrizations 
necessarily encompasses, therefore, other strategies using additional constraints, 
other than from a direct comparison with the data.     

Here we explore the constraints on the extraction of GPDs provided
by a combination of experimental data on nucleon 
form factors, PDFs, and lattice calculations of Mellin moments with $n \geq 2$. 
The latter,
parametrized in terms of Generalized Form Factors (GFFs), were  
calculated by both the QCDSF \cite{QCDSF_0,QCDSF_1,QCDSF_2,zan_talk,schi_talk} 
and LHPC \cite{LHPC_1,LHPC_2,LHPC_3}
collaborations for both the unpolarized and polarized 
cases up to $n=3$,
\footnote{In principle, lattice calculations can extend to $n=4$. This has not
been done so far because of statistical noise.}
therefore allowing to access the skewness dependence of GPDs.  
We implemented results from 
Refs.\cite{zan_talk,schi_talk} using two flavors of ${\mathcal O}(a)$-improved 
dynamical fermions. Lattice QCD calculations are increasingly improving: new 
results can, and will be added to our determination as they become available.


In order to estimate GPDs by knowing a finite set of Mellin moments we adopt  
a strategy similar to the ones originally developed in QCD based analyses 
of DIS data where a {\it mathematically rigorous} method for inverting 
moments was devised \cite{Yndurain}.  
The early QCD studies were motivated by the simplifications that Mellin moments offered in 
treating PQCD evolution. In our case, moments provided by lattice QCD represent, in addition
an important source of information on GPDs. We envisage  
a future scenario in which the calculated
Mellin moments will be embedded in a ``global'' analysis including determinations
of GPDs from experiment, and a consistent treatment of PQCD evolution. 
%

The problem of the construction of a function from its moments is mathematically
meaningful so long as one establishes 
the necessary and sufficient conditions on the moments that a unique solution exists.
A point-by-point reconstruction is therefore unattainable, however 
practical methods can be applied to a number of physics problems
including QCD based analyses of the DIS structure function data  
\cite{Yndurain} (see also \cite{Penni} for a review of early analyses), so long
as one aims at obtaining $F_2$ in a finite interval around $x_{Bj}$. 
It should be noticed, in fact, that experimental measurements 
always provide discrete ``binnings'' of kinematical variables.  
As shown in Refs.\cite{Yndurain} the Bernstein polynomials are ideal 
for reproducing the deep inelastic structure functions in that they are zero at the
endpoints, they are normalized to one, and they are peaked in different regions
within the interval $x_{Bj} \in [0,1]$. Because of the latter property the Bernstein
polynomials allow one to emphasize the behavior of the structure function at given
specific regions of $x_{Bj}$, while suppressing the others.  
It was found that $n \geq 8$ moments were necessary to give a fully quantitative
description of the behavior of $F_2(x_{Bj},Q^2)$. 
At present, only $n=3$ Mellin moments are available
from lattice QCD calculations, therefore one cannot reach a similar level of accuracy
as in the inclusive case.
In this paper we show that we can, however, reduce the kinematical region where 
information comes strictly from this type of reconstruction using Mellin
moments solely to $X<\zeta$. The parametrization we obtained in the $\zeta=0$
case \cite{AHLT1} can in fact be extrapolated smoothly to $X>\zeta$ 
(the so-called DGLAP \cite{DGLAP} region) 
by a simple 
kinematical shift in the quarks longitudinal momenta, being based on the
same partonic picture.    
The $X<\zeta$ -- ERBL \cite{ERBL} -- region does not lend itself however to 
a clearcut interpretation in terms of quark and gluon structure. 
It is in fact becoming increasingly manifest that in order to extract 
and interpret the amplitudes from a variety of hard exclusive 
experiments including DVCS in the ERBL region, 
it is fundamental to understand and connect the partonic structure 
of $t$-channel exchanges to previous information from Regge phenomenology \cite{Szc,GuzPol,Laget}.  
The contribution of   
a Regge-motivated term was already considered in quantitative studies 
at $\zeta=0$ \cite{AHLT1,VandH,DieKro}, following a similar logics as  
in the early ansatze for PDF parametrizations (see {\it e.g.} \cite{BarPhi}). 
Regge type contributions are expected to play a much more important role for GPDs at $\zeta \neq 0$, 
specifically in the  $X <\zeta$ domain,
due to the dominance of the scattering from a $q \overline{q}$ pair emerging from the 
initial nucleon in these processes (Fig.\ref{fig1}b).  

The large theoretical uncertainty in the  
$X <\zeta$ domain motivated us to perform our analysis concentrating on a 
determination of GPDs in this region, 
dictated only by experimental data and lattice moments, and therefore
in principle model independent.
The price that one pays for maintaining model independence is, however, in this
case, and with the present accuracy of lattice QCD determinations, a loss in predictivity 
in the point-by-point functional dependence of the GPDs in the ERBL region. Our results however
provide points from ab initio calculations, with given error bars in a region that cannot 
be extracted directly from experiment. These can be used in turn for guidance to constrain the 
shape of possible functional forms of GPDs, a question addressed in a forthcoming manuscript 
\cite{pap3}. Finally, improvements in the accuracy of the extracted points
will be made as more accurate lattice QCD results become available.   
 
An important consideration should also be given to the analytical properties of GPDs at $X=\zeta$.
This point marks the transition between the partonic and mesonic interpretations of GPDs, respectively,
and it acquires a particular significance only in the presence of soft matrix elements describing
the non-perturbative structure of hadrons (in {\it e.g.} a QED based description
it no longer signifies a specific breakpoint).
Simple consistency arguments related to factorization \cite{Die_rev} 
require that continuity is observed at the breakpoint. 
Recent precise measurements from Jefferson Lab indeed find non-vanishing
values of GPDs of the measured Single Spin Asymmetry (SSA) at $X=\zeta$ \cite{halla}. 
The continuity condition constitutes an additional constraint in our approach
in addition to the ones provided by the experimental data and the lattice results. 

It should be remarked that our results are specifically for the flavor Non-Singlet (NS) 
sector. 
Perturbative evolution is therefore slower than for the singlet and gluon cases,  
and it can be implemented straightforwardly similarly to the PDF case \cite{MusRad,GolMar}. 

The main thrust of the
present paper is to present an extension of the analysis of Ref.\cite{AHLT1} 
using available information at $\zeta\neq 0$.
Our method can be seen as a first step in a "bottom up" approach as opposed to other 
models presently available ({\it e.g.} the dual model of Ref.\cite{GuzPol}, and the Double Distribution 
based model of Ref.\cite{VGG}). Our aim is to make use of all information available 
both experimentally, and from ab initio calculations, and to study how to construct the functional 
dependence of the GPDs consequentially. As in similar phenomenological studies (see {\it e.g.} 
\cite{DieKro,VandH,Ansetal})
a simple and adequately flexible functional form given by the spectator diquark model 
is used in the partonic-process-dominated 
DGLAP region, with the goal of eventually easing into a global analysis. 
As for the ERBL region, our scope is to define what constraints are available, 
that come directly from ab-initio calculations. 
A similar exploratory study has not been conducted so far to our knowledge. 
The study of a functional form in X in the ERBL region that is consistent 
with the constraint found in our first quantitative study is the subject 
of a forthcoming manuscript \cite{pap3} where the real part of the 
DVCS interference term will be presented. 
As a first step, we preferred limiting the present manuscript to providing a conistent method that uses 
all information obtained from
experimental constraints plus lattice QCD 
results.
Future extensions to 
the sea quarks and gluon distributions will also help disentangle the intricate 
interplay between perturbative evolution and shape of the initial distributions. 

Our paper is organized as follows: In Section II we define the ingredients of 
our physically motivated parametrization in the $\zeta >0$ case, and introduce 
lattice moments with a prescription to account for their chiral extrapolation. 
In Section III we describe the extraction from lattice moments using Bernstein polynomials;
Our results are compared to the recent data from Jefferson Lab \cite{halla}. 
In Section IV we draw conclusions and give an outlook. 

\section{Constraints on Generalized Parton Distributions from Lattice Moments}
\subsection{Generalities}
\begin{figure}
\includegraphics[width=10.cm]{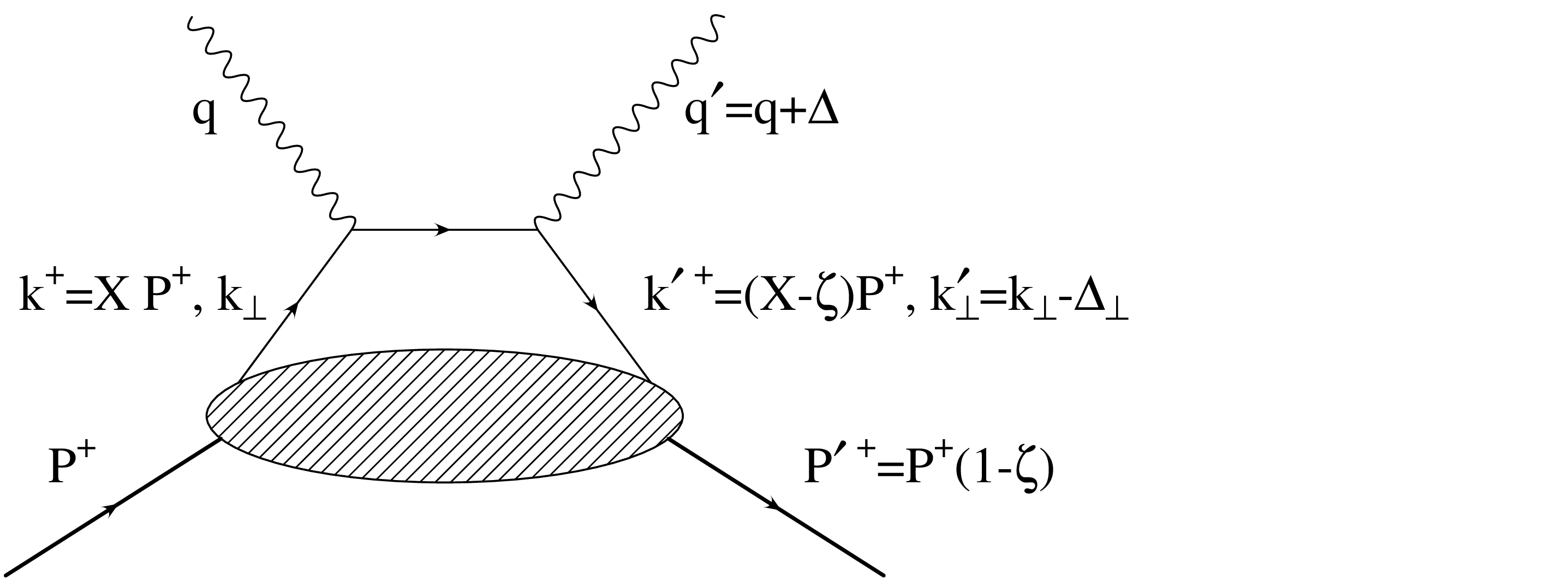}
\includegraphics[width=5.cm]{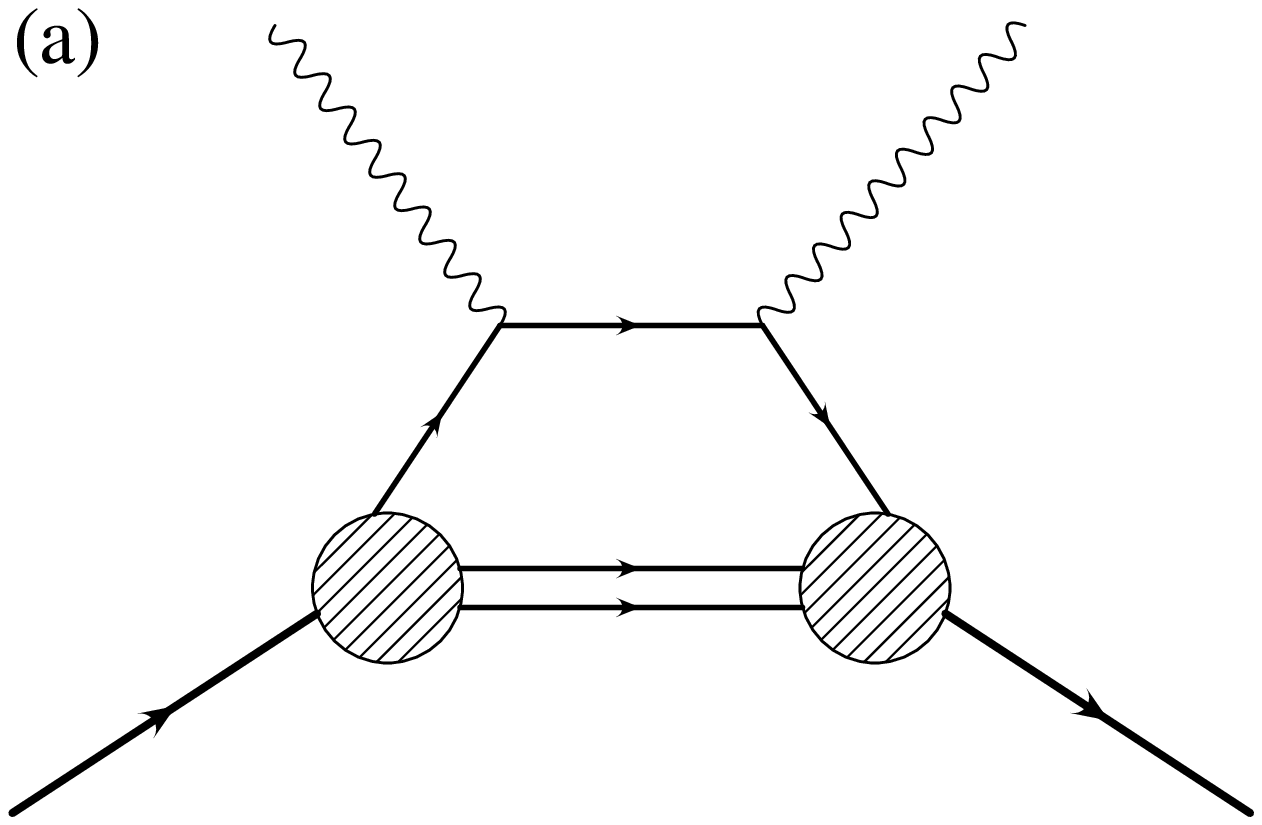}
\hspace{3cm}
\includegraphics[width=6cm]{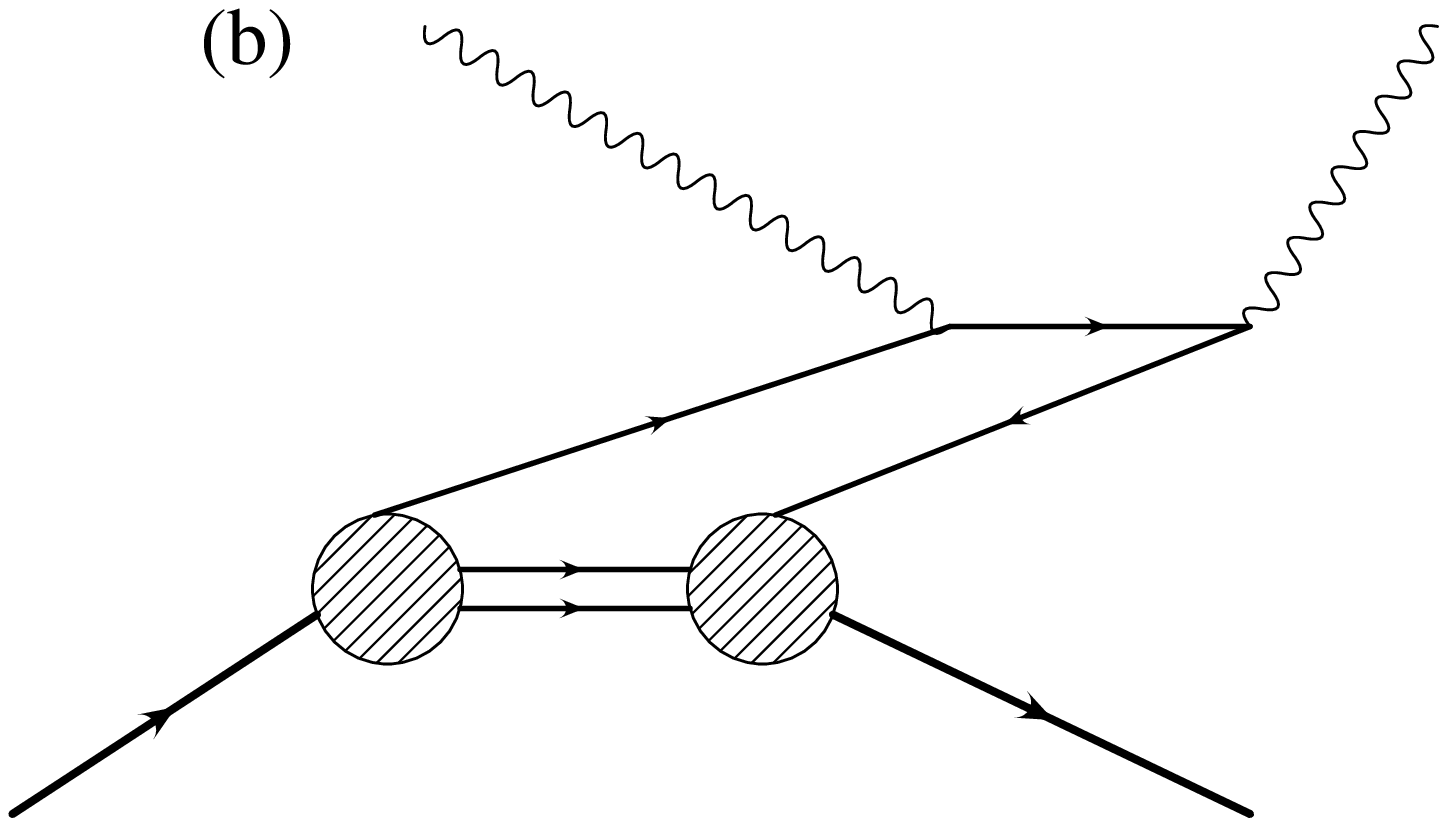}
\caption{Left: Amplitude for DVCS at leading order in $Q^2$. The light cone
components of the momenta for the active 
quarks and nucleons are explicitly written; 
Right: Time ordered diagrams for DVCS: {\bf (a)} dominant contribution in $X>\zeta$ region;
{\bf (b)} a $q \overline{q}$ pair is first produced from the nucleon and subsequently 
interacts with the photons. This process dominates the $X<\zeta$ region.
The crossed-terms where 
two of the particles in the same class are switched, are not shown in the figure.}
\label{fig1}
\end{figure}

GPDs can be extracted most cleanly from Deeply Virtual Compton Scattering (DVCS)  
at four-momentum transfer, $Q^2$, in the few/multi-GeV region 
(Fig.\ref{fig1}).   
For unpolarized scattering one has two
independent functions: $H$, and $E$, from the vector ($\gamma_\mu$)
and tensor ($\sigma_{\mu\nu}$) interactions, respectively. 
These are parametrized in terms of three kinematical invariants, besides
the initial photon's virtuality, $Q^2$. In this paper we adopt the following set:
$(\zeta, X, t)$, where $\zeta= Q^2/2(Pq)$
is the longitudinal momentum transfer between the initial 
and final protons ($\zeta \approx x_{Bj}$ in the asymptotic limit,
with Bjorken $x_{Bj} = Q^2/2M\nu$, $\nu$ being
the energy transfer associated to $\gamma^*$),
$X=(kq)/(Pq)$ is the momentum fraction relative to the initial proton carried by the struck 
parton with momentum $k$, $t = \Delta^2$, is the four-momentum transfer squared. 
In principle, the observables for DVCS and similar exclusive processes  
are functions of only two independent invariants:
$t$ and $s$ ($s=Q^2(1/x_{Bj} -1) + M^2$). 
\footnote{Throughout the paper we use the ``asymmetric'' notation with skewness parameter 
$\zeta= Q^2/2(Pq)$. 
This is  related at leading order to the  skewness parameter, $\xi$, in ``symmetric'' 
notation
by: $\xi = \zeta/(2-\zeta)$. In this representation $X=(x+\xi)/(1+\xi)$.}      
A partonic picture is however more conveniently described in terms of Light Cone (LC) 
momentum projections such as  
$\zeta$ and $X$, despite the fact that $X$ is not directly observable, 
and it appears in the amplitude as an integration variable
\cite{Die_rev,BelRad}. The need to deal with a more complicated phase space,  
in addition to the fact that DVCS interferes coherently with the Bethe-Heitler (BH) process,  
are in essence the reasons why it is more challenging to extract GPDs from experiment, 
wherefore guidance from phenomenologically motivated parametrizations becomes important. 

Initial experiments were conducted at HERMES \cite{hermes}. However, 
more recently, the high precision measurements  
from Jefferson Lab \cite{halla} have provided both the real and imaginary
parts of the Bethe-Heitler BH-DVCS interference term at  $Q^2 \approx 2$ GeV$^2$,
and $x_{Bj}=0.36$, 
showing the dominance of the twist-2 contribution. 
This, in turn, can be written as \cite{BelKirMuel_N}: 
\begin{equation}
\label{C}
\mathcal{C}^{\mathcal{I}} = F_1(t) \mathcal{H}(\zeta,t) + \frac{\zeta}{2-\zeta} G_M(t) 
\widetilde{\mathcal{H}}(\zeta,t) - \frac{t}{4M^2} 
F_2(t)\mathcal{E}(\zeta,t).
\end{equation} 
In Eq.(\ref{C}), 
$\mathcal{F} \equiv \{\mathcal{H},\mathcal{E},\widetilde{\mathcal{H}} \}$, $q=u,d,s$,
define the generalized Compton Form Factors (CFFs) for the unpolarized and polarized case, respectively;
$F_{1(2)}$ are the Dirac (Pauli) form 
factors, $G_M =F_1+F_2$   
(for details on the photon electro-production cross section and the
harmonic decomposition of its (BH)$^2$, (DVCS)$^2$, and BH-DVCS interference terms 
we refer to Refs.\cite{BelKirMuel_N,Die_rev}).

Several definitions have been given connecting the CFFs to GPDs (see the review in \cite{Die_rev}). 
We follow the notation and definitions of Ref.\cite{BroDieHwa} according to
which the leading order CFFs can be written as:
\begin{eqnarray}
\mathcal{F}(\zeta,t) = -i \pi \sum_q
e^2_q \left[F^q(\zeta,\zeta,t) - F^{q}(-\zeta,\zeta,t) \right] +
\nonumber \\ 
\mathcal{P} \int_{1-\zeta}^1 dX \left(\frac{1}{X-\zeta} + \frac{1}{X} \right) F^q(X,\zeta,t).
\label{cal_F}
\end{eqnarray}
Notice that Eq.(\ref{cal_F}) is completely analogous to the definition in the symmetric scheme 
of \cite{Ji1}, except for the choice of kinematics. 
For practical calculations it is usfeul to consider the separation into valence and sea quarks 
GPD components. Specifically,  
GPDs are related to scattering amplitudes, at variance with the PDFs, and their quark/anti-quark 
content is a function of the longitudinal momentum transfer,or skewness, $\zeta$. 
At non-zero skewness different definitions were given in the literature ({\it e.g.} \cite{GolMar,KirMuel}). 
The essence of the argument is that for GPDs three separate kinematical regions define the
scattering process:
 
\noindent 
{\it i)} $\zeta<X<1$, where the initial quark in Fig.\ref{fig1}
has momentum fraction $X$ and the final one $X-\zeta$; 

\noindent
{\it ii)} $1-\zeta<X<0$  where one scatters 
an initial anti-quark with momentum fraction $-(X-\zeta)>0$, into a final one with $-X>0$. This 
corresponds to a $u$-channel exchange (the lower limit of 
integration in Eq.(\ref{cal_F}) is defined accordingly);

\noindent 
{\it iii)} $0<X<\zeta$ where a quark-antiquark pair emerges from the nucleon and 
undergoes the electromagnetic interaction. Barring contributions from very low $X$  
($X \lesssim 0.01$) dominated by anti-quarks in the initial state, in this region the
quark carries momentum $X>0$, and the anti-quark $-(X-\zeta)>0$.

In summary, the amplitudes for scattering from either a quark or an anti-quark are defined in different
kinematical regions. As for the  identification of the valence and sea quarks contributions, 
analogously to the PDFs \cite{KutWei}, (by omitting
the $t$-dependence that is not relevant here), one has for the forward ($\zeta=0$) case 
\begin{subequations}
\label{F_zeta0}
\begin{eqnarray}
F_q(X,0) & = & q(X)  \\
F_{\bar{q}}(-X,0) & = & - \bar{q}(X), 
\end{eqnarray}
\end{subequations}
$q(X)$, and $\bar{q}(X)$ being the quark and anti-quark parton distributions from DIS.
Therefore, the valence forward GPD, $F_q^V(X,0)$, is
\begin{eqnarray}
\label{FV_zeta0}
F_q^V(X,0)  = F_q(X,0) + F_{\bar{q}}(-X,0) \equiv q(X) - \bar{q}(X),
\end{eqnarray}
and, following Ref.\cite{KutWei}, the sea quarks distribution coincides
with the anti-quarks one, $F_q^S(X,0) = \bar{q}(X)$.  
At $\zeta >0$, 
\begin{subequations}
\label{F_zetan0}
\begin{eqnarray}  
F_q(X,\zeta) & = & F_q^V(X,\zeta) + F_q^S(X,\zeta)  \; \; \; X \geq 0 \\
F_{\bar{q}}(X,\zeta) & = & F_{\bar{q}}^S(X,\zeta)  \; \; \; X \geq 0, 
\end{eqnarray}
\end{subequations}
with $-F_{\bar{q}}(-X,\zeta) = F_q(X,\zeta)$ for $-1 + \zeta \leq X < 0$.
Once the physical meaning of the various regions in $X$ is clarified, it is therefore a matter of choice 
whether or not to represent the different valence and sea quarks components of GPDs on the positive side
of the $X$ axis. 
Eqs.(\ref{F_zeta0},\ref{FV_zeta0},\ref{F_zetan0})  are in line with  Ref.\cite{KirMuel}, where 
definitions where given in the context of DVCS from nuclei, that avoid 
the appearance of spurious 
symmetries in the $X$ dependence of the off forward distributions \cite{GolMar}.
Notice the consequences of the valence and sea quarks separation 
on the structure of Eq.(\ref{cal_F}): while the first term includes a sum
over all $q$ and $\bar{q}$ distributions, the second term is given by a principal value integral
where the valence and sea quarks appear in the different domains, {\it i), ii), iii)}, explained above. 
In particular, since the valence contribution is zero at $X<0$, this implies that the $1/X$ 
term in the intergral corresponds to a singularity that cannot be handled by the integration (a similar
condition appears for $x=-\xi$ in the symmetric scheme). 
As a consequence, either the valence contribution to the CFF cannot be calculated alone, 
or the condition above dictates the dependence of the valence contribution to the GPD at $X \rightarrow 0$, 
which should be steep enough to counter the $1/X$ type singularity. We explore
this point in a forthcoming manuscript \cite{pap3}.    
Finally, as observed in the forward case, $F_q^S(X,\zeta) \neq F_{\bar{q}}^S(X,\zeta)$, in general. 
This is an aspect to be still explored {\it e.g.} for scattering from strange quarks in the exclusive channels.


We also notice that a contamination from the pure DVCS term might be present in the 
extraction of the coefficient that might be large, although weighted by a kinematical factor of 
about $1\%$ at Jlab kinematics. We give our estimate of this term in future work.  
Models such as \cite{VGG} 
do not seem to accurately reproduce the data on either the 
imaginary or real part of Eq.(\ref{C}). 
Although future experiments are planned that will allow one to 
directly determine whether such discrepancies could be also 
due to the presence of a pure DVCS contribution 
so far disregarded in the extraction of ${\cal C}$, it is at present 
important to provide 
GPD parametrizations using constraints from both inclusive deep inelastic 
and elastic scattering data, and {\it ab initio}
lattice calculations. It is our aim to explore, in what follows, the usage 
of such constraints towards a parametrization built from a bottom-up approach.

\subsection{Method Description}
In a previous publication \cite{AHLT1} we presented our parametrization of unpolarized
GPDs in the flavor Non Singlet (NS) sector based on a diquark spectator model
improved by a Regge term at low $X$. 
Parameters were given for skweness $\zeta=0$, by using constraints 
from both form factors and PDFs.    
The spectator model is ideal in this context because despite its simplicity, it 
has proven to be sufficiently flexible to 
describe (and predict) the main features of a number of distribution and fragmentation
functions in the intermediate and large $X$ regions, as well as the unintegrated PDFs  
\cite{Jakob:1997wg,Metz,Schle}.  
At $X>\zeta$,    
the proton splits into a quark carrying a LC momentum fraction $X=k^+/P^+$, transverse momentum
${\bf k}_\perp$, and a spectator
system with $1-X,  = k_X^+/P^+$, and $-{\bf k}_\perp$. After undergoing the electromagnetic
interaction, the final quark with momentum fraction $X-\zeta$, and 
the spectator system coalesce into an outgoing proton (all particles are moving forward). 
The partonic configurations are therefore equivalent to the $\zeta=0$ case, modulo a kinematical
shift. The parametrization's form is therefore (see \cite{AHLT1}):

\[ H(X,\zeta,t) = G(X,\zeta,t) R(X,\zeta,t), \]
where $G(X,\zeta,t)$ is the diquark model component, and $R(X,\zeta,t)$ the Regge-based term. 
The parametrization differs from the one in \cite{AHLT1} only by a shift to non-zero
skewness, in the kinematical 
variables (a similar form is obtained for $E(X,\zeta,t)$).

Notice the difference with other parametrizations reproducing 
the $\zeta=0$ behavior \cite{VandH,DieKro}, where the forward contribution is 
given directly by existing PDF parametrizations, $q(X)$. The latter are factored out
from so-called ``profile'' of the parton distributions, namely: 

\[ H(X,0,t) = q(X) \exp[-t f(X)].  \]
   
\noindent 
While in this case the forward limit, $H(X,0,0) \equiv q(X)$, is automatically met by 
introducing explicitly $q(X)$,
in our case, this needs to be enforced non trivially.

In other words, in Ref.\cite{AHLT1} we had to simultaneously obtain 
new parametrizations for both the GPDs and their forward limit, the PDFs at a 
low initial scale. This effort granted us, however,
the possibility to automatically extend our parametrization to  $\zeta \neq 0$ ($X>\zeta$) by a simple
variation in the kinematics. 
More details will be given in Section \ref{berns-sec}.

We subsequently analyzed the additional, $\zeta$-dependent constraints given by the higher moments of GPDs. 
The $n=1,2,3$ moments of the NS combinations: $H^{u-d} = H^u-H^d$, and $E^{u-d} = E^u-E^d$ 
are available
from lattice QCD \cite{zan_talk,schi_talk}, $n=1$ corresponding to the nucleon 
form factors.
We use such constraints within a reconstruction procedure for the GPDs from their moments,
using Bernstein polynomials. Such a procedure 
allows us to extract values for $H^{u-d}$ and $E^{u-d}$ at three values 
of $X$, defined as $\overline{X}_{k,2}$, $k=0,1,2$, for each given $\zeta$ and $t$. 
The Bernstein polynomials are used as weighting functions for 
$H^{u-d}$ and $E^{u-d}$, emphasizing  the 
regions around  $\overline{X}_{k,2}$ \cite{Yndurain,Penni}. 
As we will show in detail in Section \ref{berns-sec}, moments of GPDs calculated 
using Bernstein polynomials provide the values of $H^{u-d}$ and $E^{u-d}$,
defined as $\overline{H}^{u-d}_k$ 
and $\overline{E}^{u-d}_k$, at the points $\overline{X}_{k,2}$, $k=0,1,2$. 
Numerical results including the errors and dispersions in $X$, respectively, 
are also provided in Section \ref{berns-sec}. 

Finally, we observe that the lattice results on GPDs moments have to be 
chirally extrapolated. 
For this purpose we extended 
to the $n=2,3$ moments
a simple ansatze proposed in \cite{Ash} for the nucleon form factors. 
The procedure is illustrated in Section \ref{lattice-sec}.

Because only the first three moments are known to date, by performing our analysis in the 
whole region of $X \in [0,1]$ one obtains large theoretical uncertainties in the 
evaluated $X$ and $H, E$ values. 

We therefore determine the behavior of the GPDs at 
$X \geq \zeta$, 
by keeping the same parameter values obtained 
using the form factors and PDFs constraints in Ref.\cite{AHLT1}. 
We checked as explained later on that the curves at $X>\zeta$ are consistent with 
the values obtained with the reconstruction procedure in the whole $X$ interval.

The Bernstein polynomials reconstruction is subsequently used in a reduced region, 
the ``unknown'' $X < \zeta$, ERBL \cite{ERBL} region, where one expects a definite 
departure from 
the partonic type description of the DGLAP region.
As we shall see in Section \ref{berns-sec}, 
reducing the size of the interval in $X$ provides an advantage in   
the GPDs extraction. 

Below we summarize all the information that was used in our analysis:
\begin{itemize}
\item Accurate experimental information on the nucleon form factors. 
\item Parametrizations of the NS part of PDFs ({\it e.g.} the Alekhin set from Ref.\cite{Alekhin})
reproducing the DIS data.
\item Lattice results for the $n=1,2,3$ moments at $-t \leq 2$ GeV$^2$ with the following features:
\begin{itemize}
\item Only non-singlet contributions, $H^{u-d}$, $E^{u-d}$ are provided 
(see Section \protect\ref{lattice-sec}). 
\item The contribution from the term $C^{u-d}_{2}$, (see Eqs.(\ref{Dirac2},\ref{Pauli2}) below), 
is set equal to zero, consistently with lattice calculations
\cite{LHPC_1,LHPC_2,QCDSF_1,QCDSF_2}). 
\item The dependence on $\zeta$ is extracted 
from the term $\propto A_{32}$ (see Eq.(\ref{A3}) below). 
\end{itemize}
\end{itemize}
It should be noticed that experimental data on DVCS can 
in principle be systematically implemented in 
our extraction and will be considered in a forthcoming manuscript. 
%
{\it Predictions} for the data in 
Ref.\cite{halla} are given in Section \ref{berns-sec}.

We conclude by underlining once more that our goal is
to provide a practical method for extracting GPDs 
deriving from techniques that have been well tested in DIS. The proposed
method uses information from both experimental data and
lattice results, and it is exact 
in that for each kinematics, or $(\zeta,t)$ values, it provides 
the values of $H(X,\zeta,t)$ and $E(X,\zeta,t)$ 
in at given $X$ values, with calculable theoretical error. 
With the proposed approach we also wish to provide 
an alternative based on a bottom-up type of analysis, at variance 
with top-down models proposed so far including the 
Double Distribution (DD) hypothesis \cite{VGG},
on the Mellin-Barnes integral representation \cite{KumMuel}, and/or on  
the dual representation of Ref.\cite{GuzPol}.
Our approach, similarly to what found for DIS \cite{Yndurain} provides 
a different perspective on the problem of a formal extraction of GPDs from their moments 
that tends to be more subject to uncontrolled numerical ambiguities such as the 
one in {\it e.g.} the oscillating term 
inherent in the integral defining the continuation to complex $n$.

\subsection{Extrapolation of Mellin Moments from Lattice QCD}
\label{lattice-sec}
The Mellin moments of GPDs are most clearly described in the symmetric frame of \cite{Ji1},
where the relevant kinematical variables are $x=(k^{\prime \, +} + k^+)/(P^{\prime \, +} + P^+) 
\equiv ((X-\zeta/2)/(1+\zeta/2)$ and 
$\xi=-2 \Delta^+/(P^{\prime \, +} + P^+) \equiv \zeta/(2-\zeta)$.
The $x$ moments of GPDs are defined as
\begin{eqnarray}
H_{n}^q(\xi,t) & = & \int_1^1 dx x^{n-1} H^q(x,\xi,t) \\
E_{n}^q(\xi,t) & = & \int_1^1 dx x^{n-1} E^q(x,\xi,t),  
\end{eqnarray}
where $q=u,d,s$ and we disregarded strange quarks contributions.
$H_{n}^q$ and $E_{n}^q$ represent the form factors of local twist two operators.
From the 
Lorentz structure of these form factors, one obtains the following polynomiality relations in $\xi$, 
made explicit by using the Gordon decomposition: 
\begin{eqnarray}
\label{polyH}
H_{n}^{q}(\xi,t)
& = & \sum_{i=0}^{\frac{n-1}{2}}A^{q}_{n,2i}(t) \xi^{2i} + {\rm mod}(n,2) \xi^{n} C^q_{n}(t)
\\
\label{polyE}
E_{n}^{q}(\zeta,t)
& = & \sum_{i=0}^{\frac{n-1}{2}}B^{q}_{n,2i}(t) \xi^{2i} - {\rm mod}(n,2) \xi^{n} C^q_{n}(t).
\end{eqnarray}
$A^q_{n,2i}(t)$, $B^q_{n,2i}(t)$, and $C^q_{n,2i}(t)$ are 
the Generalized Form Factors (GFFs).
The latter were calculated on the lattice by both the LHPC and the QCDSF groups for 
the unpolarized, polarized and transversely polarized cases 
\cite{LHPC_1,LHPC_2,LHPC_3,QCDSF_0,QCDSF_1,QCDSF_2}. 
Since the GFFs correspond to off-diagonal matrix elements of twist-two operators, 
it was necessary to introduce a new technique implementing all $H(4)$ cubic group 
operators and index combinations producing the same continuum GFFs \cite{LHPC_1} 
in order to provide statistically 
accurate lattice measurements from an overdetermined set of equations.  
Therefore, lattice calculations extend
only up to $n \leq 3$. We list below the GFFs expressions for $n=1,2,3$  using the notation 
of \cite{Ji1,LHPC_1}. 

For $n=1$ one obtains the nucleon Dirac and Pauli form factors: 
\begin{eqnarray}
\label{Dirac}
H_{1}^q & \equiv  & A_{10}^q(t) = F_1^q(t) = \int\limits_{-1}^1 dx H^q(x,\xi,t) = \int\limits_{-1+\zeta}^1 \frac{dX}{1-\frac{\zeta}{2}} H^q(X,\zeta,t)  \\
\label{Pauli}
E_{1}^q & \equiv & B_{10}^q(t) = F_2^q(t) = \int\limits_{-1}^1 dx E^q(x,\xi,t) = \int\limits_{-1+\zeta}^1 \frac{dX}{1-\frac{\zeta}{2}} E^q(X,\zeta,t).  
\end{eqnarray}
For $n=2$ one has:
\begin{eqnarray}
\label{Dirac2}
H_{2}^{q} & = & A_{20}^q(t) + \left(- \frac{2\zeta}{2-\zeta} \right )^2 C_{2}^q(t)
\\
\label{Pauli2}
E_{2}^{q} & = & B_{20}^q(t) - \left(- \frac{2\zeta}{2-\zeta} \right )^2 C_{2}^q(t)
\end{eqnarray}
In the limit $t \rightarrow 0$,  Eqs.(\ref{Dirac},\ref{Pauli}) give the 
baryon number, $B_q$, and anomalous magnetic moment, $\kappa_q$:
\begin{eqnarray}
\label{charge}
A_{10}^q(0) & \equiv & B_q =  \int\limits_0^1 dX  q(X) 
\\
\label{anom_mag}
B_{10}^q(0) & \equiv & \kappa_q  
\end{eqnarray}
Eq.(\ref{Dirac2}) gives the LC momentum fraction 
carried by quark $q$:
\begin{eqnarray}
A_{20}^q(0) & \equiv & \langle x \rangle_q =  \int\limits_0^1 dX X q(X) 
\label{momSR}
\end{eqnarray}
Furthermore, Eqs.(\ref{Dirac2},\ref{Pauli2}) are   
related through the angular momentum sum rule \cite{Ji1}: 
\begin{eqnarray}
A_{20}^q(0) + B_{20}^q(0) =  2 J^q,
\end{eqnarray}
where $J^q= L^q + S^q$ is the total -- orbital plus intrinsic -- angular momentum
carried by the quark $q$. 

Finally, the $n=3$ moments are given by:
\begin{eqnarray}
\label{A3}
H_{3}^{q} & = & A_{30}^q(t) + \xi^2  A_{32}^q(t)
\\
\label{B3}
E_{3}^{q} & = & B_{30}^q(t) + \xi^2  B_{32}^q(t).
\end{eqnarray}

In our calculation we use lattice calculations for the unpolarized GFFs for $n=1,2,3$ 
obtained by the QCDSF collaboration using two flavors of ${\mathcal O}(a)$-improved 
dynamical fermions for several values of $t$
in the interval 
$0 \lesssim t \lesssim 5$ GeV$^2$, and covering a range of pion mass values, 
$m_\pi \gtrsim 500 \, {\rm MeV}^2$ \cite{zan_talk,schi_talk}.
Similarly to previous evaluations \cite{LHPC_1} 
the GFFs for both $H$ and $E$ display a dipole type behavior for all three $n$ values,
the value of the dipole mass increasing with $n$. 

A straightforward implementation of present lattice calculations 
in realistic parametrizations is hampered by
the rather large discrepancy with the experimental data, 
associated to the large 
values of the pion mass, $m_\pi$ used in Refs.\cite{QCDSF_1,QCDSF_2,LHPC_1,LHPC_2}.  
Early extrapolations used a linear approximation \cite{schi_talk} that although 
improving the comparison with experiment, is not sufficient to 
grant the accuracy that is necessary for a quantitative 
parametrization. 
Extrapolations using Chiral Perturbation Theory (ChPT) of the GPD moments up to $n=2$ 
are currently being addressed using different methods: Heavy Baryon ChPT (HBChPT), in 
Refs.\cite{DieMan,Dor}, self-consistently improved ChPT \cite{Beane}, and finite-range regulators 
techniques \cite{Mat,PWang}. 
While most results have bean focused on the values at $t=0$, which are important
for the determination of the nucleon orbital angular momentum,  
equations for $t \neq 0$ were given explicitely in \cite{Dor}. 
Whether this can be applied to the large $m_\pi$ masses where most lattice results 
are provided, is still a matter of intense debate that is beyond the scope of this paper.
In addition, the number of 
parameters of the calculation is rather large to enable a really precise evaluation 
of the moments at $t \neq 0$. Nevertheless evaluations providing the ``t'' term with a  
$\approx 25\%$ error have been 
possible within the most recent set of lattice results \cite{Marina_Philipp_priv}.         
On the other side, the finite-regulator technique results by allowing for an extension of
the range in pion mass considered, will conisderably reduce the uncertainty on the $n=2$ moment 
determination,   
While a number of new extrapolation methods are currently being explored 
we adopted a simple ansatz for the extrapolation of the dipole masses 
for the Dirac and Pauli form factors \cite{Ash} which: {\it i)} uses the connection between
the dipole mass and the nucleons radius; {\it ii)} introduces a modification of the non analytic 
terms i
the standard chiral extrapolation that suppresses the contribution of chiral loops at large $m_\pi$.
Despite its simplicity the conjecture of \cite{Ash} well reproduces the trend of 
lattice results at large $m_\pi$ 
while satisfying the main physical criteria {\it i)} and {\it ii)}.
The resulting
values for the ``physical'' nucleon dipole masses are both sensibly different 
than using a linear extrapolation, and much closer to the experimental data. 
As a consequence, the proton
and neutron magnetic form factors, as well as the proton electric form factors 
can be reproduced 
quite accurately, whereas non-negligible discrepancies were found 
only in the case of the more elusive neutron form factor. 
Based on the good agreement with experiment obtained using this extrapolation method,
we extended it to the recent lattice data on $n=2,3$ GPD moments \cite{zan_talk}.
We reiterate that the choice of this simple method in spite of recent ChPT developments
is justified in the context of this work by the fact that on one side we need all 
available higher moments, up to $n=3$, and these are not accessible with available methods; 
secondly, the recent flurry of activities has been focused on the $t=0$ region, while we 
are primarily interested in $t\neq 0$ where the uncertainties in the various ChPT-based 
determinations become more important.
Finally, the goal of our work is to suggest an approach to determine GPDs using all available
constraints from both experiment and theory. While hoping that both present and future 
efforts will provide us with increasingly precise results, and with a method to determine 
larger $n$ moments, the role of this work is limited to implementing them within our suggested
analysis as a theoretical input with a precisely determined uncertainty, independently 
from the merit of any specific approach used to obtain them in the first place.     
A detailed study of different sources of theoretical uncertainties is beyond the scope
of this work, and will be performed in a forthcoming paper.   

The chiral behavior of the form factors is exhibited by the isovector and isoscalar contributions:
\begin{eqnarray}
\label{isovector}
G_{M(E)}^V & = & G_{M(E)}^p - G_{M(E)}^n \\
\label{isoscalr}
G_{M(E)}^S & = & G_{M(E)}^p + G_{M(E)}^n,
\end{eqnarray}
where $G_{M(E)}^{p(n)}$ are the magnetic (electric) form factors for the proton (neutron).

Lattice results are given only for the isovector components, related to to the 
corresponding form factors as
\begin{eqnarray}
\label{vec_magn}
G_M^V & = & (F_1^p - F_1^n) + (F_2^p - F_2^n) \\ \nonumber 
& = & \int \, dX (H^u - H^d) + \int \, dX (E^u - E^d)
\end{eqnarray}
and
\begin{eqnarray}
\label{vec_elec}
G_E^V & = & (F_1^p - F_1^n) - \tau (F_2^p - F_2^n) \\ \nonumber 
& = & \int \, dX (H^u - H^d) - \tau \int \, dX (E^u - E^d),
\end{eqnarray}
with $\tau=-t/4M^2>0$. From Eqs.(\ref{vec_magn}) and (\ref{vec_elec}) one obtains
\begin{eqnarray}
\label{HuHd}
H_1^{u-d} \equiv \int \, dX (H^u - H^d) & = &  \frac{\tau G_M^V + G_E^V}{1+\tau} \\ 
E_1^{u-d} \equiv \int \, dX (E^u - E^d) & = & \frac{G_M^V - G_E^V}{1+ \tau}.
\end{eqnarray}

For $\zeta =0$ it was found that the
$n=2,3$ moments from Ref.\cite{LHPC_1,QCDSF_1} display a dipole behavior:
\begin{equation}
(H_{M(E)}^V)_n(t) = \frac{H_{M(E)}^V(0)}{ \left( 1+(-t)/(\Lambda_{M(E)}^{V, \, n})^2 \right)^2 },  
\end{equation}  
where in this case:
\begin{eqnarray}
\label{HuHdn}
H_n^{u-d} \equiv \int \, dX X^{n} (H^u - H^d) & = & \frac{\tau (H_M^V)_n + (H_E^V)_n}{1+\tau} \\
\label{EuEdn}
E_n^{u-d} \equiv \int \, dX X^{n} (E^u - E^d) & = &  \frac{(E_M^V)_n - (E_E^V)_n}{1+ \tau},
\end{eqnarray}
where the l.h.s. quantities are obtained from the lattice moments 
calculations, whereas $(H_{M(E)}^V)_n$ and $(E_{M(E)}^V)_n$ are amenable to chiral 
extrapolations. 

We surmise that for $n=2,3$, similarly to the $n=1$ case, a relation of inverse proportionality 
exists between the dipole mass and a 
given hadronic length scale, 
similar to the following relation 
between the magnetic and electric dipole masses, $\Lambda_M^V$ and $\Lambda_E^V$, and the 
corresponding mean squared radii exists, namely: 
\begin{eqnarray}
\langle r^2 \rangle_{M(E)}^V & = & \frac{12}{(\Lambda_{M(E)}^V)^2} ,
\end{eqnarray}   
and that a chiral behavior for this hadronic scale
can in principle be formulated. Lacking an exact expression for the 
coefficients in such an extrapolation, we performed
a fit using the following form:   
\begin{eqnarray}
(\Lambda_{M(E)}^{V \,, n})^2  & = & \frac{12(1 + \alpha_n^{M(E)} m_\pi^2)}{\beta_n^{M(E)} + 
\gamma_n \ln \left( \frac{m_\pi^2}{m_\pi^2 + \mu^2} \right)}
\label{lambda_second} 
\end{eqnarray}   
In Eq.(\ref{lambda_second}) $\alpha_n^{M(E)}$, $\beta_n^{M(E)}$ and $\gamma_n$ 
are parameters to be fitted to the lattice results; $\mu = 0.4-0.5$ GeV is a ``cut-off'' 
parameter for the pion mass, above which chiral loops are suppressed. 

Eq.(\ref{lambda_second}) differs from the form factor expression from Ref.\cite{Ash} in that
we let the coefficient of the non linear term, $\gamma_n$, be a parameter of the equation
whereas in the $n=1$ case, an exact expression for this term 
was obtained in Ref.\cite{Meis}, namely: $\gamma_1 = g_A^2 M_N/(8 \pi f_\pi^2 \kappa_V)$, 
$g_A = 1.26$ being the axial 
coupling constant, and $f_\pi = 93$ MeV being the pion decay constant. 
\footnote{
Note that a term $\propto \arctan(\mu/\pi)$, present in \protect\cite{Ash} 
was dropped for simplicity from the isovector magnetic
term. The effect of disregarding this term  for the first moment 
was evaluated to be within the errors from the lattice calculation.}

\begin{figure}
\includegraphics[width=12.cm]{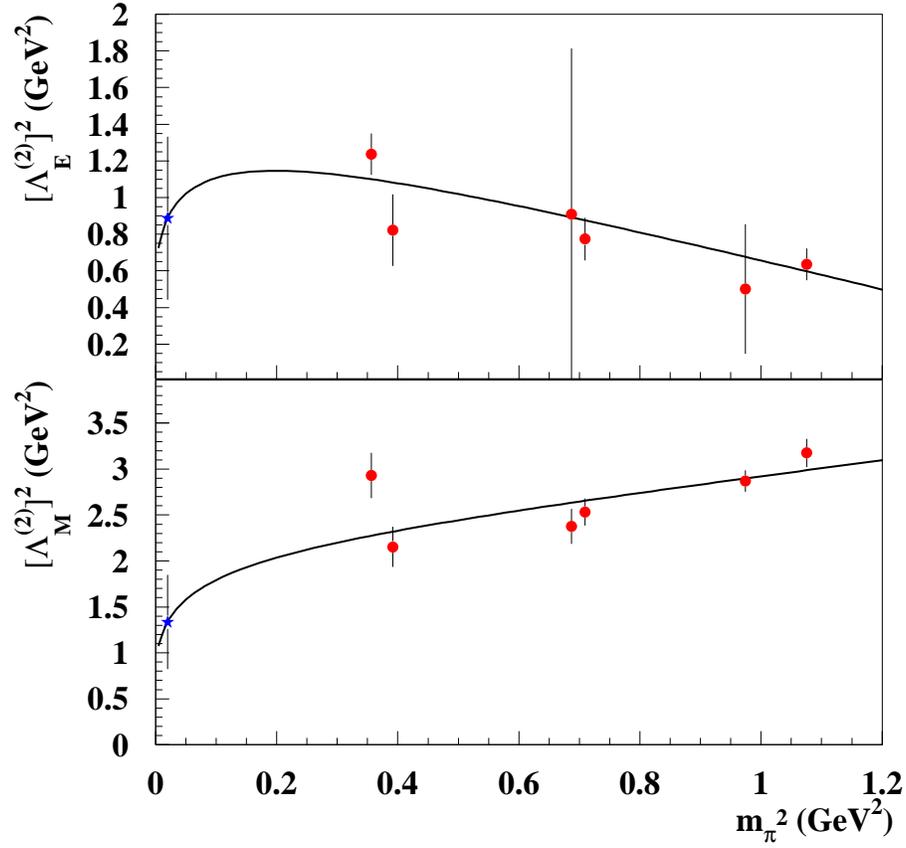}
\caption{(color online) The dipole masses squared for $n=2$, for the isovector magnetic (lower panel) 
and electric (upper panel) contributions obtained by performing fits to the 
lattice results of \protect\cite{zan_talk} (full circles). The lines represent the result of 
our fitting procedure using the ansatz in Eq.(\ref{lambda_second}). The value at the physical
pion mass obtained from our fit is also shown (star).} 
\label{fig2}
\end{figure}

 
Eqs.(\ref{HuHdn}), (\ref{EuEdn}) were first fitted using a dipole form for
$A_{n0}^{u-d}(t) \equiv A_{n0}^{u}(t) - A_{n0}^{d}(t)$ and 
$B_{n0}^{u-d}(t)\equiv B_{n0}^{u}(t) - B_{n0}^{d}(t)$, $n=1,2,3$ \cite{QCDSF_1},
for all provided sets of lattice results.
For $n=1$ the lattice results are consistent with what was obtained in Ref.\cite{Ash}.
The values obtained at the physical pion mass 
are given in Table \ref{table_lambda}. 
The dipole masses at $n=2$, $(\Lambda_{E(M)}^{V \, 2})^2$, are shown in Fig.\ref{fig2}
along with 
our results for the fits using Eq.(\ref{lambda_second}), with $\mu=0.4$ GeV, in accordance with 
\cite{Ash}.

\begin{figure}
\includegraphics[width=12.cm]{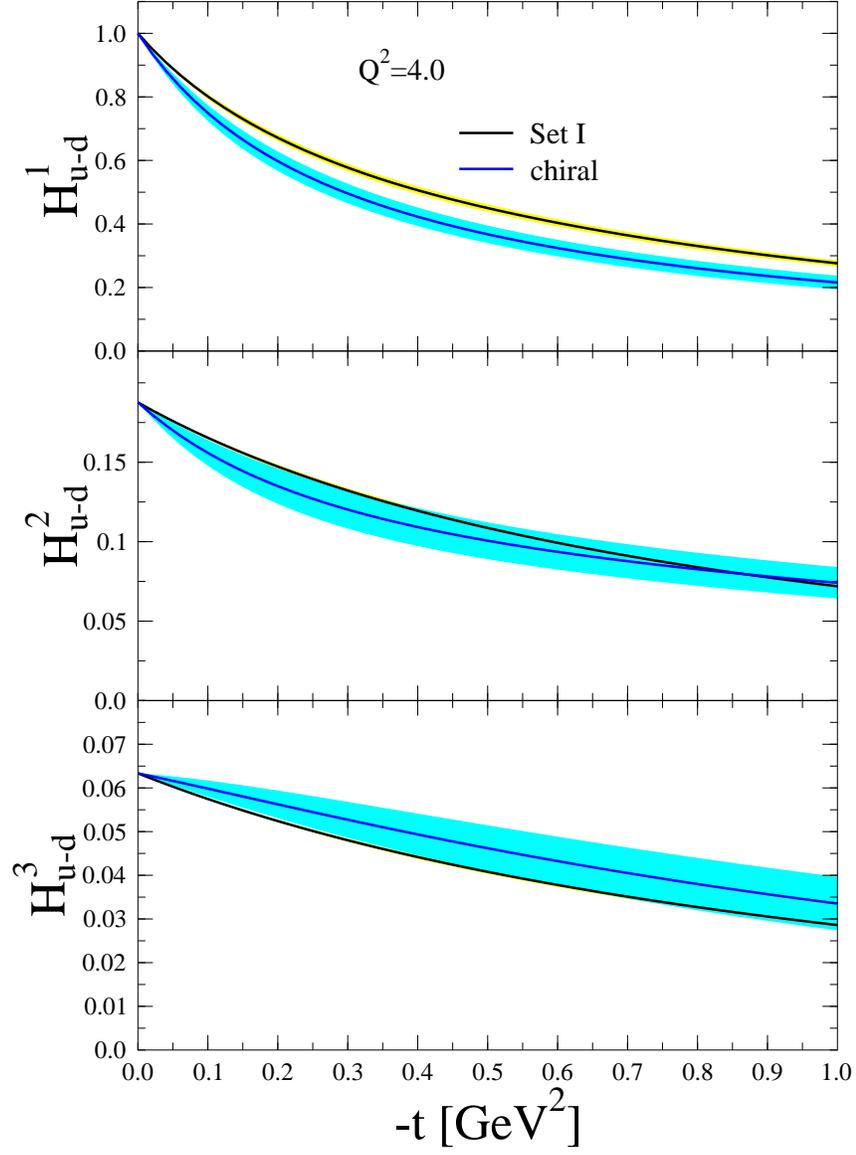}
\caption{(color on-line) Moments $A_{n0}^{u-d}(t)$ plotted vs. $-t$, 
for $n=1,2,3$, obtained with the
values from Table \ref{table_lambda}, compared to the parametrization from Ref.\cite{AHLT1}.
For $n=1$ (upper panel) the parametrization for the form factor from Ref.\cite{AHLT1} 
is also shown along with the theoretical error band.} 
\label{fig4}
\end{figure}

\begin{table}
\center
\begin{tabular}{|c|c|c|}
\hline
$n$  & $(\Lambda_E^V)^2$ (GeV$^2$) &  $(\Lambda_M^V)^2$ (GeV$^2$) \\ \hline
$1$  &  $0.457 \pm 0.048$ &  $0.576 \pm 0.060$ \\ \hline
$2$  &  $0.704 \pm 0.163$ & $1.371 \pm 0.238$  \\ \hline
$3$  &  $1.80 \pm  0.57$ & $ 1.25 \pm 0.43 $ \\ 
\hline
\hline
\end{tabular}
\caption{Values of the dipole masses squared, $(\Lambda_E^V)^2$ and $(\Lambda_M^V)^2$, for the vector 
electric and vector magnetic moments, for $n=1,2,3$, as obtained from the fit of the
lattice results described in the text.}
\label{table_lambda}
\end{table}
Fig.\ref{fig4} shows the moments $A_{n0}^{u-d}(t)$, for $n=1,2,3$, obtained with the
values from Table \ref{table_lambda}. For $n=1$ the extrapolated lattice
results are compared to the parametrization from Ref.\cite{AHLT1} that 
accurately reproduces the experimental data for the electric and magnetic 
form factors. Although there is a discrepancy with the parametrization/experimental
data, its size is consistent with the previous evaluation from Ref.\cite{Ash}.  
The error band on the lattice evaluations is a result of the fit 
error on the dipole mass.
Finally, we also extracted the value for $A_{32}(t) = 0.0302 \pm 0.0103$ (Eq.(\ref{A3})), 
by performing a linear fit of the lattice results. 

We also note that similar results can be obtained for 
both $E$ and the polarized GPDs, and  will be presented in a forthcoming paper \cite{pap3}. 
In the next Section, in order to illustrate our method, we concentrate on the function $H$, which 
displays the highest accuracy, being constrained in the forward limit. 

\section{Reconstruction from Bernstein Polynomials}
\label{berns-sec}
With a finite number of moments in hand, one can apply to GPDs
a reconstruction method 
following the approach initially used for PDFs in Ref.\cite{Yndurain}.
The procedure, that uses only moments with integer $n$, introduces 
a complete set of positive definite polynomials, 
the Bernstein polynomials, from which one constructs weighted averages of the 
GPDs, around definite values of $X$.   
As pointed out in \cite{Yndurain}, because of inherent uncertainties in the 
experimental binnings, one does not need a point-wise description of the
structure functions/GPDs, but {\it averages} of these quantities over 
ranges of $X$, thus making the method both mathematically sound and appealing
from the practical point of view.  
The method is in fact an alternative 
to a reconstruction of  the GPDs from their 
inverse Mellin transforms that however requires in principle the knowledge of the behavior with $n$ of 
the Mellin moments. 
Anyhow, such approaches were used for a quantitative reconstruction of PDFs 
from the deep inelastic scattering data \cite{BurGae}; it was more recently 
used in NNLO analyses in \cite{MochVer}, and they were also suggested in order
to evaluate GPDs at $\zeta=0$ \cite{schi_talk}.
In a more general case of $\zeta$-dependent GPDs, 
merging information from both experimental data and lattice, and taking into account 
PQCD evolution, clearly involves a more complex phase space than for PDF studies, 
and makes the implementation
of the inverse Mellin transform method a prohibitive task.  

\begin{figure}
\includegraphics[width=12.cm]{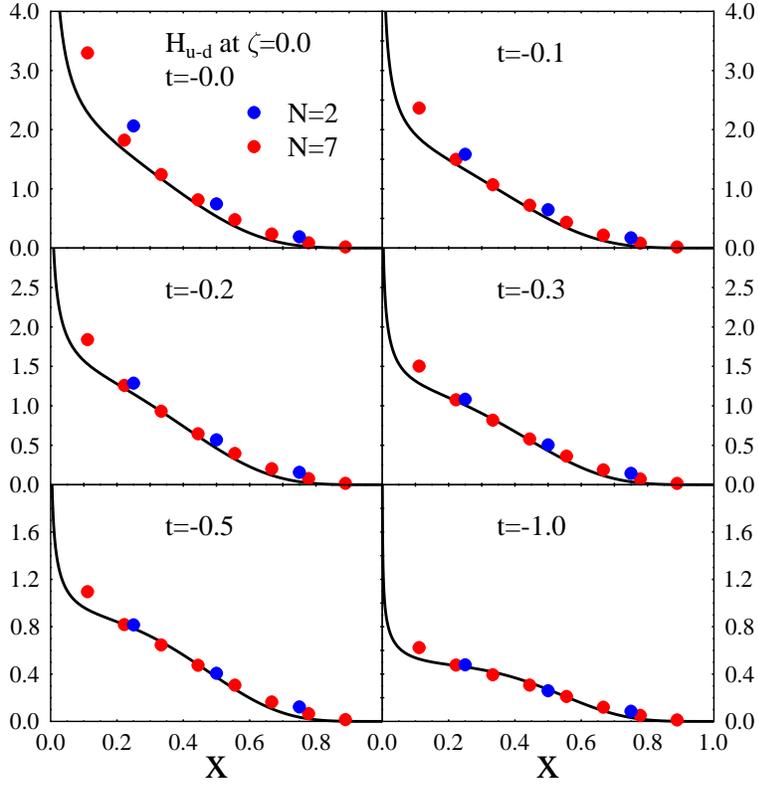}
\caption{(color on-line) Comparison between $H^{u-d}$ calculated at $\zeta =0$, for
different values of $t$ ($0\leq -t \leq 1$ GeV$^2$), using the parametrization
in Ref.\protect\cite{AHLT1} (full curve), and the reconstructed function using 
the first eight, $n=7$, and three, $n=2$, Bernstein moments.} 
\label{fig5}
\end{figure}

We consider therefore:
\begin{equation}
\overline{H}_{k,n}(\zeta,t) = \int\limits_0^1 H(X,\zeta,t) \, b_{k,n}(X) dX \; \; \; \; k=0,...n, 
\label{berns1}
\end{equation}
where the weight functions are the Bernstein polynomials:
\begin{equation}
b_{k,n}(X) = \frac{X^k \, (1-X)^{n-k}}{\int\limits_0^1 X^k \, (1-X)^{n-k} dX },
\end{equation} 
that, by definition, are functions that are concentrated on restricted ranges 
within the $X \in [0,1]$ interval, specifically around the values:
\begin{equation}
\overline{X}_{k,n} = \int\limits_0^1   b_{k,n}(X) \, X dX = \frac{k+1}{n+2},
\end{equation}
with a width given by the dispersion $\Delta_{k,n}$ \cite{Yndurain,Penni}
\footnote{Notice that the dispersion in Eq.(\ref{width}) is different from the 
definition used in Ref.\cite{Max}}.
\begin{eqnarray}
\label{width}
\Delta_{k,n}^2 & = &  \overline{X^2}_{k,n}  - \overline{X}_{k,n}{\, ^2} 
= \frac{(k+1)(n-k+1)}{(n+2)^2 (n+3)}
\end{eqnarray}
By writing $b_{k,n}$ in Eq.(\ref{berns1}) with a binomial expansion, one obtains:
\begin{equation}
\overline{H}_{k,n}(\zeta,t) = \frac{(n+1)!}{k!} \sum_{l=0}^{n-k} \frac{(-1)^l}{l! (n-k-l)!} H_{l+k+1}
\end{equation}
where $H_{l+k}$ are the Mellin moments:
\begin{equation}
H_{l+k+1} = \int\limits_0^1 H(X,\zeta,t) \, X^{l+k} \, dX.
\end{equation}
We tested the range of validity of the proposed reconstruction approach 
by using an already known function, namely our parametrization for 
$H(X,\zeta,t)$ at $\zeta=0$ 
\cite{AHLT1}. 
In Fig.\ref{fig5} we show $H(X,0,t)$ along with a reconstruction using 
the first eight $(n=7)$ and three $(n=2)$ 
Bernstein moments, respectively. From the figure it is clear
that it is sufficient to consider eight  moments to accomplish an 
accurate description
of $H^{u-d}$, at $X \gtrsim 0.1$, and for all values of $t$. With only three moments, slightly
larger discrepancies arise. It should be also noticed that the Reconstruction using Bernstein
polynomials is always less accurate at small $X$, a feature already noticed in \cite{Yndurain}.
Such a discrepancy is however small with respect to current error bars on the lattice 
moments determinations. Moreover, if necessary, {\it i.e.} 
as more accurate lattice results and data become available, 
it can be improved upon either by implementing 
higher order corrections \cite{Yndurain}, or by using modified Bernstein 
functions as {\it e.g.} in \cite{Max}.      

\begin{figure}
\includegraphics[width=12.cm]{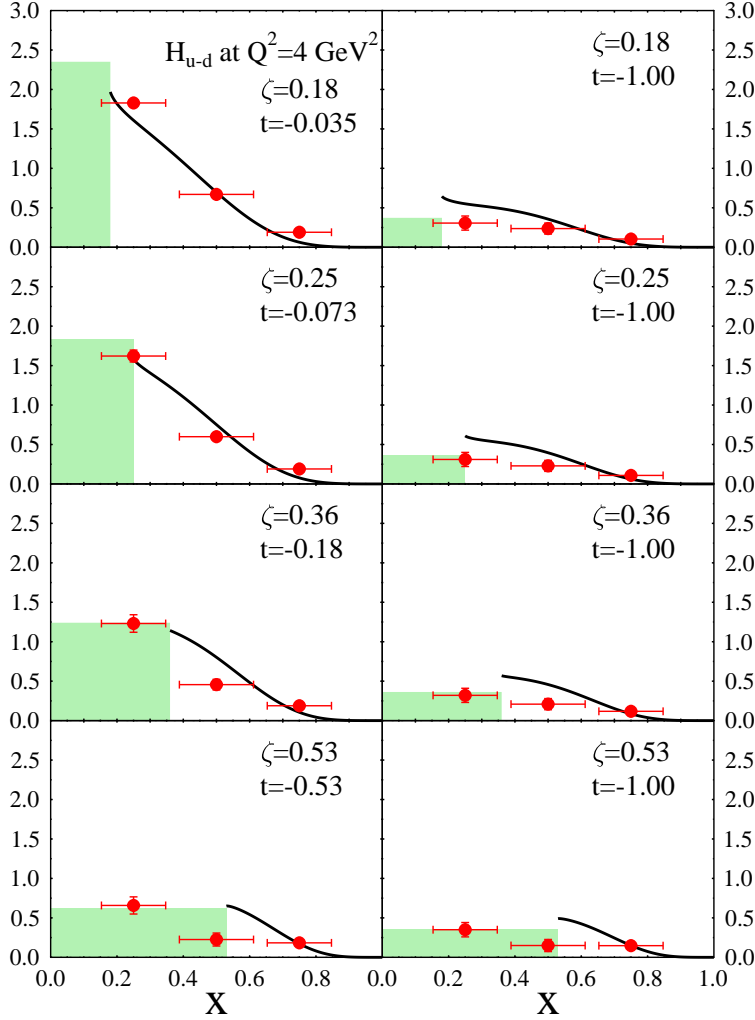}
\caption{(color on-line) Comparison of $H^{u-d}$ for 
different values of $\zeta = 0.18, 0.25, 0.36, 0.53$, 
$-t \equiv t_{min}= 0.035, 0.073, 0.18, 0.53$ GeV$^2$ (left panel), and $-t = -1$ GeV$^2$ (right panel), 
calculated using the parametrization from Eqs.(\ref{param1_H}), (\ref{diq_zeta}), (\ref{regge_zeta}) 
(full curves), and the Bernstein polynomials based reconstruction using  
the first three Bernstein moments. The latter were calculated from the chirally extrapolated
lattice QCD moments (Fig.\ref{fig4}), at the scale $Q^2= 4$ GeV$^2$. The horizontal error bars 
are obtained
by evaluating the dispersion, Eq.(\ref{width}). The error bars on the moments are mostly not visible
on the plot. The shaded area represents the ``missing area'' in the ERBL region, obtained imposing
the polynomiality condition (see text).} 
\label{fig6}
\end{figure}

\begin{figure}
\includegraphics[width=16.cm]{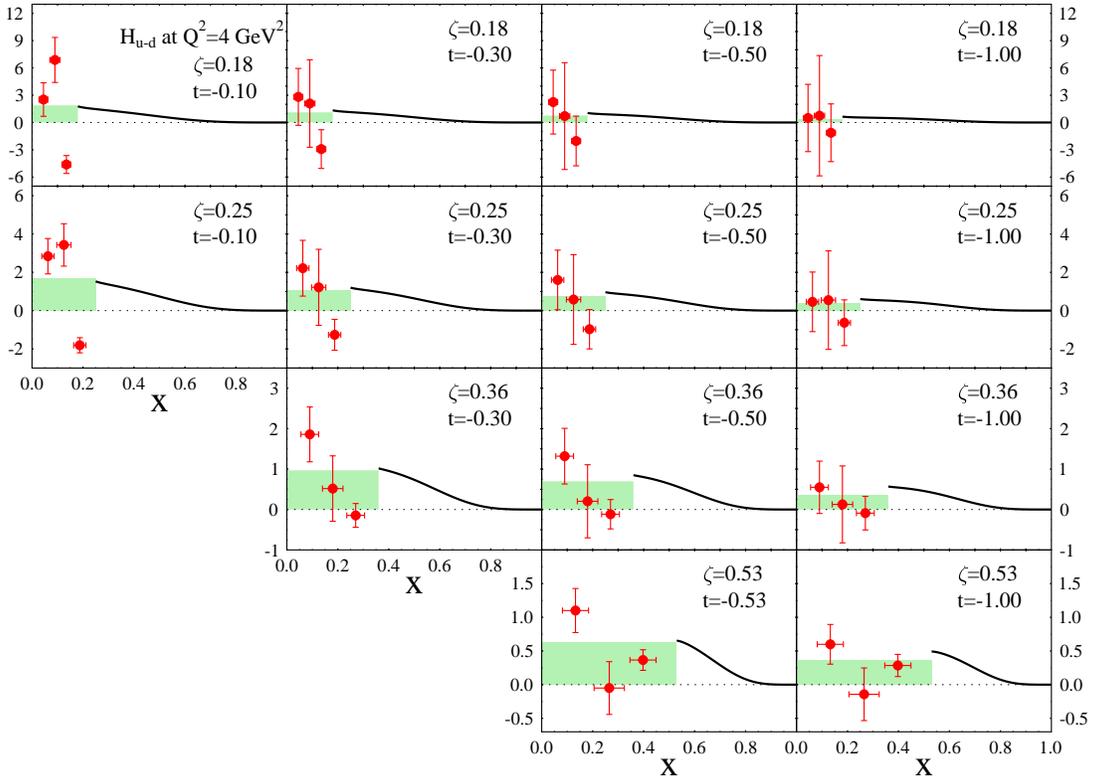}
\caption{(color on-line) $H^{u-d}$ in the ERBL region obtained with the method
described in the text, for 
different values of $\zeta = 0.18, 0.25, 0.36, 0.53$, and 
$0.10 \leq -t \leq 1$ GeV$^2$. } 
\label{fig7}
\end{figure}

In Figures \ref{fig6} and \ref{fig7} we use the three available lattice moments 
(Section \ref{lattice-sec}) to reconstruct $H^{u-d}$. 
In this case, for each value of $\zeta$ and $t$, 
one obtains the following values of $\overline{X}$: 

\[ \overline{X}_{02} = 0.25, \; \overline{X}_{12}=0.5, \; \overline{X}_{22} =0.75, \]
and weighted average values of $\overline{H}$:
\begin{subequations}
\begin{eqnarray}
\label{berns3_x1}
\overline{H}_{02}( \overline{X}_{02})& = & 3A_{10} - 6 A_{20} + 3 
\left[A_{30} + \left( -\frac{2\zeta}{2-\zeta} \right)^2 A_{32} \right],
\\
\overline{H}_{12}(\overline{X}_{12}) & = & 6 A_{20} - 6 
\left[ A_{30} +  \left( -\frac{2\zeta}{2-\zeta} \right)^2 A_{32} \right],
\\   
\overline{H}_{22}(\overline{X}_{22}) & = & 3 \left[ A_{30}+ 
 \left( -\frac{2\zeta}{2-\zeta} \right)^2 A_{32} \right]. 
\end{eqnarray}
\end{subequations}
In Fig.\ref{fig6} we show the results for the weighted averages, with their given width calculated 
according to Eq.(\ref{width}). The full curve in the DGLAP region  
was obtained by extending to $\zeta \neq 0$ the parametrization 
from Set I of Ref.\cite{AHLT1}
as follows:
\begin{subequations}
\label{HandE}
\begin{eqnarray}
H^I(X,\zeta,t) & = & G_{M_{X}^I}^{\lambda^I}(X,\zeta,t) \, R^I_1(X,\zeta,t) 
\label{param1_H}
\\
E^I(X,\zeta,t) & = & \kappa \, G_{M_X^I}^{\lambda^I}(X,\zeta,t) \, R^I_2(X,\zeta,t)
\label{param1_E}
\end{eqnarray}
\end{subequations}
where $G_{M_{X}^I} ^{\lambda^I}$ is given by a spectator model
including both scalar and axial-vector components for the final diquark:
\begin{equation}
\label{diq_zeta}
G_{M_{X}^I}^{\lambda^I}(X,\zeta,t) = 
{\cal N} \frac{X}{1-X} \int d^2{\bf k}_\perp \frac{\phi(k^2,\lambda)}{D(X,{\bf k}_\perp)}
\frac{\phi({k^{\prime \, 2},\lambda)}}{D(X,{ \bf k}_\perp^\prime)}.   
\end{equation}
Here $k$ and $k^\prime$ are the initial and final quark momenta respectively (Fig.\ref{fig1}a), 
$D(X,\zeta,{\bf k}_\perp) \equiv  k^2 - m^2$, $D(X,\zeta,{\bf k}_\perp^\prime) \equiv  k^{\prime \, 2} 
- m^2$, ${\bf k}_\perp^\prime ={\bf k}_\perp - (1-X)/(1-\zeta)\Delta$, $m$ being the struck quark mass, 
$\kappa$ the quark's anomalous magnetic moment, 
$\Delta=P-P^\prime$ being the four-momentum transfer, and: 
\begin{eqnarray}
k^2 & = & X M^2 - \frac{X}{1-X} M_X^{2}  - \frac{{\bf k}_\perp^2}{1-X} \\
k^{\prime \, 2} & = & \frac{X-\zeta}{1-\zeta} M^2 - 
\frac{X-\zeta}{1-X} M_X^{2} - \left({\bf k}_\perp - \frac{1-X}{1-\zeta} \Delta \right)^2 \frac{1-\zeta}{1-X},
\end{eqnarray}
with $M$, the proton mass, and $M_X$ the (flavor-dependent) diquark mass (we suppress the flavor
indices for simplicity).
Furthermore, $\phi(k^2,\lambda)$ defines the  vertex functions in both the
scalar and axial-vector cases \cite{AHLT1}.  
The normalization factor includes the nucleon-quark-diquark coupling, and it
is set to ${\cal N} = 1$ GeV$^6$. $G_{M_{X}^I}^{\lambda^I}(X,\zeta,t)$ reduces to 
the form given in Ref.\cite{AHLT1} in the $\zeta \rightarrow 0$ case.

The functions $R^I_{1(2)}$ were introduced in \cite{AHLT1} in 
order to simultaneously account for: {\it i)} the low $X$ behavior through 
an extra Regge motivated, term, $X^{-\alpha}$, which
is fundamental for obtaining the correct baryon number and momentum sum rules; 
{\it ii)} the small $t$ behavior, while preserving the asymptotic behavior 
of the form factors. For $\zeta=0$ they read:
\begin{equation}
\label{regge_0}
R^I_{1(2)} =  X^{-\alpha^I -\beta_{1(2)}^I (1-X)^{p_{1(2)}^I} t}
\end{equation}
Notice that the resulting parameters for $R^I_{1(2)}$ found in Ref.\cite{AHLT1} 
are sensibly different for $H$ and $E$, since at variance with the forward case
governed by $G_{M_{X}}^I$ where no specific constraint exists for $E$, 
these two functions are impacted on differently by 
the Dirac and Pauli form factors, respectively. 
Moreover, the fit parameters are not numerically similar to the
ones from Regge phenomenology since the diquark term defined in 
Eqs.(\ref{param1_H}), (\ref{param1_E}) keeps on being important at low $X$and $t$
thus interfering with the Regge part of the parametrization. 
The overall behavior of $H$  
is however consistent with the parameter values from Regge phenomenology (see discussion
in \cite{AHLT1} and \cite{VandH}). 
Finally, it should be remarked that Eqs.(\ref{HandE}) present an approximate treatment 
of the spin structure, and that improved spectator models such as the ones recently 
worked out in Ref.\cite{Metz} should be adopted for a full consideration of all 
GPDs, and of their spatial d.o.f. interpretation. 
This is true in our case also for the function $E$ where we chose a rather crude 
approximation, justified in our case by the goal of our paper that largely 
makes use of moments of GPDs. It should be kept in mind that this 
approximation is expected to eventually break down at very large $X$, as it 
already can be seen in Ref.\cite{AHLT1} 
in reproducing sensible values for the transverse displacements $s^q$ .
A full treatment of the spin structure of GPDs is however beyond the scope of this
paper and it will be considered elsewhere.

For $\zeta \neq 0$, $R^I_{1(2)}$ is also modified with respect to the $\zeta=0$ case
through a $\zeta$-dependent kinematical shift in $t$ resulting from the fact that 
\begin{eqnarray}
- t & = & \frac{4 \xi^2 M^2}{1-\xi^2} + (1-\xi^2) {\bf D}^2 \equiv -t_{min} + (1-\xi^2) {\bf D}^2
\end{eqnarray} 
where the Fourier conjugate to the impact parameter is now 
${\bf D} = {\bf P}^\prime/(1-\xi) - {\bf P}/(1+\xi)$, and 
$\xi  =  \zeta/(2-\zeta)$ (${\bf D}$ reduces to $\Delta$ at $\zeta=0$ \cite{Die_impact}), 
yielding: 
\begin{equation}
\label{regge_zeta}
R^I_1(X,\zeta,t) = X^{-\alpha^I -\beta_{1(2)}^I (1-X)^{p_{1(2)}^I} (t+t_{min})}
\end{equation}
More details on this transformation are given in Appendix \ref{regge_app}.
From Fig.\ref{fig6} one can notice a very good agreement between the Bernstein reconstructed 
values, and the parametrization in the DGLAP region obtained by 
perturbatively evolving at Leading Order (LO) Eqs.(\ref{param1_H}), (\ref{param1_E}) to the 
scale $Q^2=4$ GeV$^2$, where the lattice results where evaluated.
We also show the missing area (shaded area in the figure) obtained through the polynomiality
condition, by ensuring that the sum of the areas in the ERBL and DGLAP regions, respectively,
gives the form factor value at the given value of $t$.  

The Bernstein polynomials based reconstruction was subsequently repeated in the ERBL region only. 
This was accomplished by assuming 
the validity of the parametrization in the DGLAP region, and by calculating 
reduced moments in the ERBL region through:
\begin{eqnarray}
\label{missing_a}
\left( H_{n} \right)_{X<\zeta} \equiv H_{n}^\zeta = H_{n} - \left( H_{n} \right)_{X>\zeta},
\end{eqnarray}   
where $\left( H_{n} \right)_{X>\zeta}$ is given by:
\begin{eqnarray}
\left( H_{n} \right)_{X>\zeta} = \int\limits_\zeta^1 \,  H^I(X,\zeta,t) X^{n-1} \, dX  \; \; \; n=1,2,3
\end{eqnarray}   
with $H^I(X,\zeta,t)$ given by Eq.(\ref{param1_H}), while $H_{n}$ are the lattice moments. 
The procedure defined in Eqs.(\ref{berns3_x1}) was repeated for the reduced moments, 
yielding:
\begin{subequations}
\label{berns_zeta}
\begin{eqnarray}
\overline{H}_{02}( \zeta X_{02})& = & \frac{1}{\zeta^3} \left\{ 3 A_{10}^\zeta \, \zeta ^2 
- 6 A_{20}^\zeta \, \zeta  + 3 
\left[A_{30}^\zeta + \left( -\frac{2\zeta}{2-\zeta} \right)^2 A_{32} \right] \right\},
\\
\overline{H}_{12}(\zeta X_{12}) & = & \frac{1}{\zeta^3} \left\{ 6 A_{20}^\zeta \, \zeta - 6 
\left[ A_{30}^\zeta +  \left( -\frac{2\zeta}{2-\zeta} \right)^2 A_{32} \right] \right\},
\\   
\overline{H}_{22}(\zeta X_{22}) & = & \frac{1}{\zeta^3} \left\{ 3 A_{30} + 
\left( -\frac{2\zeta}{2-\zeta} \right)^2 A_{32}  \right\} , 
\end{eqnarray}
\end{subequations}
Eqs.(\ref{berns_zeta}) were obtained by redefining 
the Bernstein polynomials in the region $X \in [0,\zeta]$ as:
\begin{equation}
b_{k,n}(X,\zeta) = \frac{X^k \, (\zeta-X)^{n-k}}{\int\limits_0^\zeta X^k \, (\zeta-X)^{n-k} \, dX} ,
\end{equation} 
the dispersion (Eq.(\ref{width}), being evaluated as:
\begin{eqnarray}
\Delta_{k,n}^\zeta & = & \zeta \Delta_{k,n}. 
\end{eqnarray}
Results using the reduced moments 
are presented in Fig.\ref{fig7}, along with our evaluation in the DGLAP region using 
Eq.(\ref{param1_H}). 
The missing areas, which value is given by $A_{10}^\zeta \equiv H_1^\zeta$, are also highlighted in 
the figure.  
The theoretical errors displayed in the figure are from two sources: the lattice errors, and the 
fit errors (see Fig.\ref{fig4}), added in quadrature. 
One can see that the effect
of reducing the interval over which the weighted areas are taken, produces much larger errors 
on the moments, due to the presence of powers of $1/\zeta$ in  Eqs.(\ref{berns_zeta}). 
The determination of $H^{u-d}$ is therefore more precise at larger $\zeta$ values ($\zeta \gtrsim 0.3$).  
The property of polynomiality is satisfied by construction. 
Furthermore, we also notice that the area 
spanned by the Bernstein moments, and the value of the missing area calculated from 
Eq.(\ref{missing_a}) are in agreement within the theoretical errors.  
We reiterate that explicit model dependence is minimized in the determination of $H^{u-d}$ in the
ERBL region because of the phenomenological constraints obtained  
from the form factor data, the PDF parametrizations for the Non-Singlet (NS) sector, 
and lattice QCD results. 
Finally, the results in the ERBL region seem to display a visible pattern in the
$\zeta$, $t$, and $X$ variables. This will allow us in principle to construct a 
specific form for a parametrization driven by phenomenological constraints only, 
that will be present in future work \cite{pap3}. 

\begin{figure}
\includegraphics[width=12.cm]{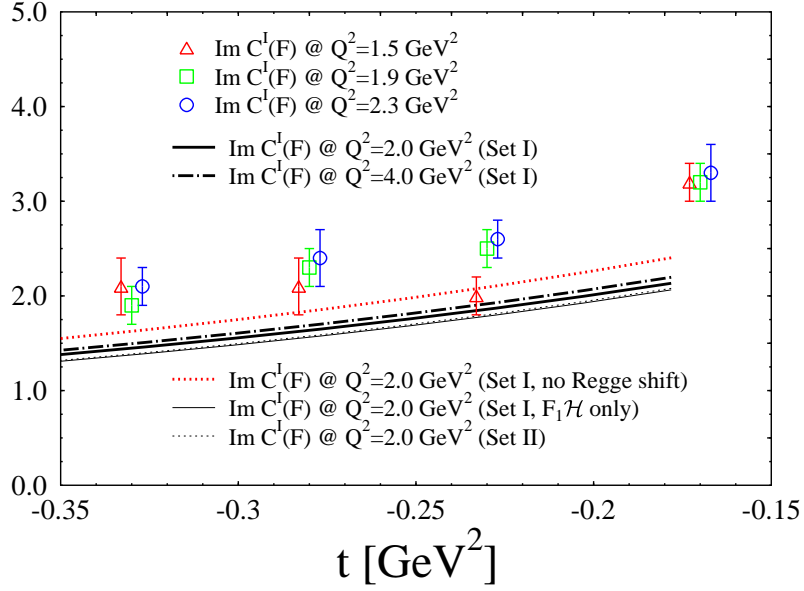}
\caption{(color on-line) The imaginary part of 
the Bethe-Heitler BH-DVCS interference term at leading order, $C^I(F)$, Eq.(\ref{C}) for a proton
target. 
Experimental data are from Ref.\cite{halla} at $x_{Bj}=0.36$ and $Q^2=1.5, 1.9, 2.3$ GeV$^2$. 
The full curve shows our prediction, including the theoretical error, evolved
to $Q^2=2$ GeV$^2$. The dot-dashed curve, obtained at $Q^2=4$ GeV$^2$, 
shows that the effect of evolution is relatively small.   
The other curves represent variations of the parametrization obtained respectively 
by disregarding the contribution from $E$, in Eq.(\ref{C}), by disregarding the 
kinematical shift in the Regge term described in the text, and by using $E$ from 
Set II of Ref.\cite{AHLT1}.} 
\label{fig9}
\end{figure}

\begin{figure}
\includegraphics[width=12.cm]{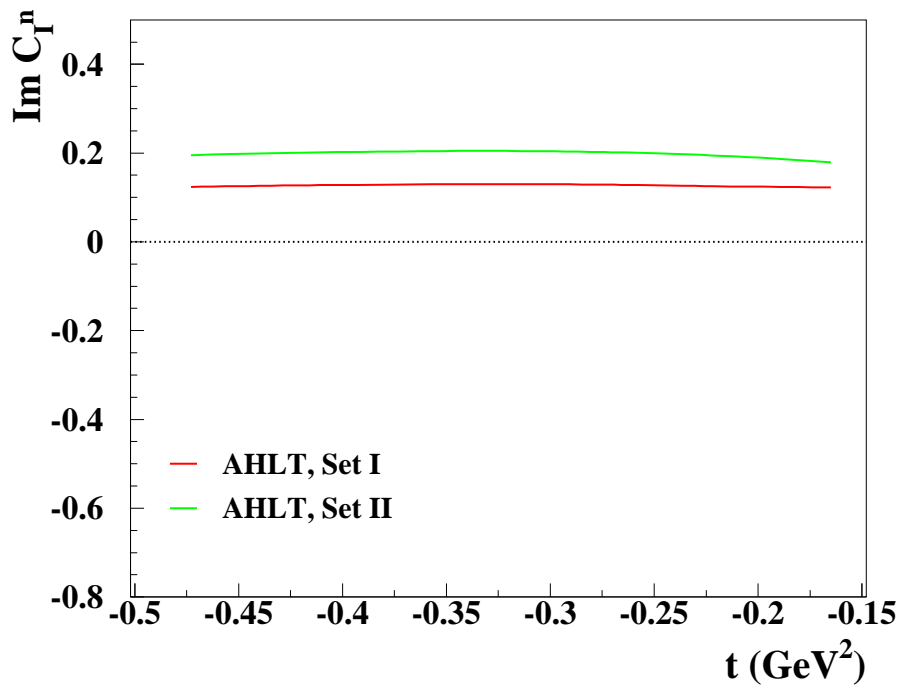}
\caption{(color on-line) The same as Fig.\protect\ref{fig9} for a neutron target. 
Calculations were performed in the kinematical range 
of Ref.\cite{Mazouz}, namely at $Q^2 \approx 2$ GeV$^2$ and $x_{Bj}=0.36$.
The two curves are variations of the parametrization using $E$ from 
Set I and Set II of Ref.\cite{AHLT1}, respectively.} 
\label{fig10}
\end{figure}

In Figs.\ref{fig9} and \ref{fig10} we show the results of our parametrization 
at $\zeta=X$ for the imaginary part of 
the Bethe-Heitler BH-DVCS interference term at leading order, Eq.(\ref{C}), for
proton and neutron, respectively.
The proton results are compared with recent data from Jefferson Lab \cite{halla}, 
at $x_{Bj}=\zeta=0.36$, and $Q^2 \approx 2$ GeV$^2$, while the neutron
ones are presented at the kinematics of the forthcoming analysis from Ref.\cite{Mazouz} .
Notice that Jefferson Lab kinematics is expected to be dominated by NS contributions and it
is therefore ideal for a comparison with the model developed here. 
We observe an excellent agreement with the data for all variants of our parametrization, 
except for perhaps the largest value in $t$, that
lies, however, at the margin of the experimental acceptance. 
While the proton contribution is governed by $H$, 
the neutron one is directly sensitive to $E$, as shown also in  our calculation 
using two variants of our fit that differ sensibly 
in the treatment of the less constrained $E$ distribution. 
We also tested the role of PQCD evolution in this region by showing results at two
different scales, $Q^2=2$ and $4$ GeV$^2$, respectively. As expected in the NS sector, 
PQCD evolution turns out to be slow, and it is therefore not to be considered a major concern.

\section{Conclusions}
We presented a parametrization of the unpolarized GPDs, $H$, and $E$, 
constructed by using a spectator model for the quark-nucleon 
scattering amplitude in which the forward limit is obtained by 
fitting directly to PDFs in the non singlet sector.   
This is at variance with exisiting phenomenologically constrained models where 
the forward limit is obtained using directly 
the results from exisiting parametrizations of PDFs, and it  
allows us to better study the 
role of Regge-type exchanges, that are disengaged, in our case, from the specific form 
used for the  parton distributions.
The other constraints defining our parametrization at zero skewness are 
provided by the electric and nucleon form factor data.
By analyzing directly the vertex structure of the scattering amplitude 
we extended our predictions to the non-zero skewness case 
obtained by modifying the kinematical variables' dependence in our expressions
for the quark scattering (DGLAP) dominated region. 
In the $X < \zeta$ (ERBL) region, where a quark-antiquark pair from the
initial proton participates in the scattering process, we took into account
additional constraints provided by the higher, $\zeta$-dependent, 
moments of GPDs. 
These are in principle available from recent lattice calculations, provided 
reasonable chiral extrapolation procedures are implemented.
A reconstruction procedure based on Bernstein polynomials, 
that allows to extract in a largely model independent way the GPDs in the
elusive ERBL region, is proposed. 
We showed that with a larger number of moments (up to $n=8$), a perspective
perhaps possible in the near future \cite{Detmold}, one could in principle
obtain a highly accurate description of GPDs in a wide portion of the phase
space.  

Differently from the double distribution based approaches 
that represent a model calculation, 
the proposed method is, for the first time to our knowledge, 
an attempt to obtain a realistic parametrization constrained by 
phenomenological input, namely by  
experimental data on form factors and PDFs in combination with lattice results. 
Given the paucity of current direct experimental measurements of GPDs, 
our goal is to provide more stringent and less model dependent
predictions that can be useful for both model builders, in order 
to gain insight on the dynamics of GPDs, and for experimenters in 
planning future DVCS and other related hard exclusive scattering experiments. 

Finally, our approach can be compared to recently developed models relying on 
t-dependent moments, namely the dual model, a project to be pursued in future investigations.
In addition, a more detailed treatment of the spectator model used in developing our
parametrization is on its way, that will allow to extend our approach to the 
spin dependent sector where many lattice results exist, and where a rich experimental
program is being pursued.  

\acknowledgments
We thank James Zanotti for providing us with the recent lattice 
calculations from the QCDSF collaboration.
We also thank Stan Brodsky, Marina Dorati, Philipp Haegler, Peter Kroll, Gerrit Schierholz, 
Tony Thomas, 
Eric Voutier, Ping Wang, and Ross Young for useful comments.  
This work is supported by the U.S. Department
of Energy grant no. DE-FG02-01ER41200. 

\appendix
\section{Extrapolation of Regge term in the DGLAP region at $\zeta \neq 0$} 
\label{regge_app}
At $\zeta \neq 0$, the contribution to the parametrization $R^I_{1(2)}$ in 
Eqs.(\ref{param1_H}, \ref{param1_E}), although 
not governed by the kinematical shifts affecting the diquark model kinematics, 
is modified with respect to the $\zeta=0$ case
presented in \cite{AHLT1} through a $\zeta$-dependent kinematical shift in $t$.
The shift results from the connection with 
the formulation of GPDs in impact parameter space \cite{Die_impact}, where one defines:
\begin{equation}
q(x,\xi,{\bf b}) = \int d^2 {\bf D} \, e^{-i {\bf D} \cdot {\bf b}} \, H(x,\xi,{\bf D}) 
\end{equation}
where ${\bf D}$, the Fourier conjugate to the impact parameter {\bf b}, is defined as 
\begin{equation}
{\bf D} = \frac{{\bf P}^\prime}{1-\xi} - \frac{{\bf P}}{1+\xi},
\end{equation} 
and: 
\begin{eqnarray}
- t & = & \frac{4 \xi^2 M^2}{1-\xi^2} + (1-\xi^2) {\bf D}^2 \equiv -t_{min} + (1-\xi^2) {\bf D}^2
\end{eqnarray} 
(${\bf D}$ reduces to $\Delta$ at $\zeta=0$ \cite{Die_impact}).
In the overlap representation:
\begin{equation}
q(x,\xi,{\bf b}) = \sqrt{1-\xi^2} \psi^*\left(\frac{x-\xi}{1-\xi}, \frac{{\bf b}}{1-\xi} \right) 
\psi\left(\frac{x+\xi}{1+\xi}, \frac{{\bf b}}{1+\xi} \right)
\end{equation}
By considering gaussian type wave functions, with parameter $\lambda$, 
one obtains by Fourier transformation, a form for the profile function 
consistent with  Eq.(\ref{regge_0}), namely:
\begin{equation}
H(x,\xi,t) \propto e^{-{\bf D}^2/A(x,\xi)} 
\end{equation}
where:
\begin{equation}
A(x,\xi) = \frac{2 \lambda^2}{1-x} \frac{1}{1-\xi^2}
\end{equation}
Therefore, for $\xi \neq 0$ one expects the modification with respect to 
the $\xi=0$ case of the variables in the profile
function to be given by:
\begin{eqnarray}
-t & \rightarrow & \left( -t - \frac{4 \xi^2 M^2}{1-\xi^2} \right) \frac{1}{1-\xi^2} \\
1-x & \rightarrow & (1-x) (1-\xi^2)
\end{eqnarray}
Keeping into account both modifications simultaneously yields the argument
of  Eq.(\ref{regge_0}): 
\begin{equation} 
-t  \rightarrow   -t - \frac{4 \xi^2 M^2}{1-\xi^2} \equiv -t + t_{min}
\end{equation}


\newpage


\end{document}